\documentclass[conference]{IEEEtran}
\IEEEoverridecommandlockouts
\usepackage{cite}
\usepackage{amsmath,amssymb,amsfonts}
\usepackage{algorithmic}
\usepackage{graphicx}
\usepackage{textcomp}
\usepackage{xcolor}
\def\BibTeX{{\rm B\kern-.05em{\sc i\kern-.025em b}\kern-.08em
    T\kern-.1667em\lower.7ex\hbox{E}\kern-.125emX}}

\usepackage{color}
\newcount\Comments  
\definecolor{darkgreen}{rgb}{0,0.5,0}
\definecolor{purple}{rgb}{1,0,1}
\newcommand{\kibitz}[2]{\ifnum\Comments=1\textcolor{#1}{#2}\fi}

\usepackage[absolute,overlay]{textpos}
\usepackage{tabularx}
\usepackage{subcaption} 
\usepackage{url}
\usepackage{xpatch}
\usepackage{lscape}
\usepackage[misc,geometry]{ifsym}

\begin{document}
\begin{textblock*}{20cm}(1cm,1cm)
\textcolor{red}{Preprint accepted by the fast abstract track of DSN2021. To appear in the proceedings of DSN2021 (supplemental volume).}
\end{textblock*}

\title{
	Detecting Operational Adversarial Examples for Reliable Deep Learning
}

\author{\IEEEauthorblockN{Xingyu Zhao, Wei Huang, Sven Schewe, Yi Dong, Xiaowei Huang}
\IEEEauthorblockA{\textit{Department of Computer Science} \\
\textit{University of Liverpool}\\
Liverpool, U.K. \\
\{xingyu.zhao,w.huang23,sven.schewe,yi.dong,xiaowei.huang\}@liverpool.ac.uk}
}

\maketitle

\begin{abstract}
The utilisation of Deep Learning (DL) raises new challenges regarding its dependability in critical applications. Sound verification and validation methods are needed to assure the safe and reliable use of DL. However, state-of-the-art debug testing methods on DL that aim at detecting adversarial examples (AEs) ignore the operational profile, which statistically depicts the software's future operational use.
This may lead to very modest effectiveness on improving the software's delivered reliability, as the testing budget is likely to be wasted on detecting AEs that are unrealistic or encountered very rarely in real-life operation.
In this paper, we first present the novel notion of ``operational AEs'' which are AEs that have relatively high chance to be seen in future operation.
Then an initial design of a new DL testing method to efficiently detect ``operational AEs'' is provided, as well as some insights on our prospective research plan.
\end{abstract}


\section{Context and Motivation}
\label{sec_contect_movtiation}

The field of Verification and Validation (V\&V) of Deep Learning (DL) software has recently enjoyed much activity and progress, with \textbf{robustness} being one of the properties, and probably \textit{the} property, in the limelight.
DL Robustness requires that the decision of the DL model is invariant against small perturbations on inputs.
While definitions of robustness vary, they share the intuition that all inputs in an input region $\eta$ have the same prediction label, where $\eta$ is usually a small norm ball (defined in some $L_{p}$-norm distance)
around an input $x$. Inside $\eta$, if an input $x'$ is classified differently to $x$ by the DL model, then $x'$ is an \textit{adversarial example}\footnote{Although named as ``adversarial'', $x'$ could be either a benign input with perturbations from natural environments or a malicious attack from attackers. We confine this research to the former, given it is hard to model and predict attackers' behaviours by a distribution over the input space.} (AE) -- a ``failure point'' (in the input space) that caused by ``bugs''. 

All V\&V methods for DL robustness are essentially about \textit{detecting} AEs.
There are emerging studies on systematically evaluating the AE detection ability of state-of-the-art V\&V methods, e.g.\ \cite{harel_canada_is_2020}.
One of the criteria is \textit{naturalness}, for which the work \cite{harel_canada_is_2020} introduces quantitative metrics to assess how \textit{realistic} the generated test cases are.
Because fixing the AEs detected by realistic/natural inputs would have more practical impact on the DL model's dependability.
Indeed, testing DL models with natural/realistic inputs is neither new nor against common sense; DeepXplore \cite{pei_deepxplore_2017}, e.g., uses domain-specific constraints to generate test cases that are valid and realistic.


However, the \textbf{delivered reliability} as a \textit{user-centred} property \cite{littlewood_software_2000} requires more than just detecting natural/realistic bugs. A vivid example in \cite{adams_optimizing_1984} shows that it would take 5,000 years of execution (based on users' day-to-day operation) to reveal about one third of the bugs in some tradition software systems.
Clearly, spending all testing budget on detecting those ``5,000 years bugs'' is unwise.
This is also true for DL software:
given a limited testing budget, to reveal as many (potentially rare) AEs as possible is misleading.  
We therefore have to focus on areas in the input space that are more likely to be executed by users:
we want to detect  AEs from the high density area of the \textit{Operational Profile} (OP) -- a probability distribution defined over the whole input domain quantifying how the software will be operated \cite{musa_operational_1993}.

Remarkably, our new ``operational AEs'' concept is  more stringent than realistic/natural AEs \cite{zhao2018generating}: operational AEs are realistic/natural, but not vice versa.
While studies on detecting realistic/natural AEs emerge, there is no dedicated techniques for detecting operational AEs, which motivates this research.

\section{Research Objectives}
\label{sec_research_obj}

Our main objective is to design and implement a new ``debug'' testing method for DL that ensures the OP information is explicitly considered, so that the detected AEs have practical impact on the delivered reliability. Specifically, compared to the state-of-the-art DL testing methods, we aim to integrate and optimise the trad-off between the following information:

\textit{a)} The OP, which is not necessarily the same as the distribution of existing data (the training and testing datasets) nor constant after deployment.

\textit{b)} Naturalness. Ideally the OP returns the probability of being executed for each point in the input space. However, practically we might not have a sound OP estimator at the very fine-grain level for every single input, rather a coarse-grain level for a ``cell'' of inputs (e.g.\ a small norm ball around a natural point input).
Thus, as a fallback solution, we have to apply quantified naturalness as an approximation to the ``local OP'' inside each cell.

\textit{c)} Gradient of Loss.
Generating test cases faithfully according to the OP (including naturalness as an approximation of the ``local OP'') is categorised as \textit{operational testing}, which is known to be inefficient in detecting bugs \cite{frankl_evaluating_1998}.
Given a limited testing budget, e.g. a number of test cases, the gradient of loss over the input space must be incorporated to fulfil our goal of detecting as many ``operational AEs'' as possible.

\section{Research Questions and Methodology}
\label{sec_research}

\begin{figure}[h!]
	\centering
	\includegraphics[width=0.48\textwidth]{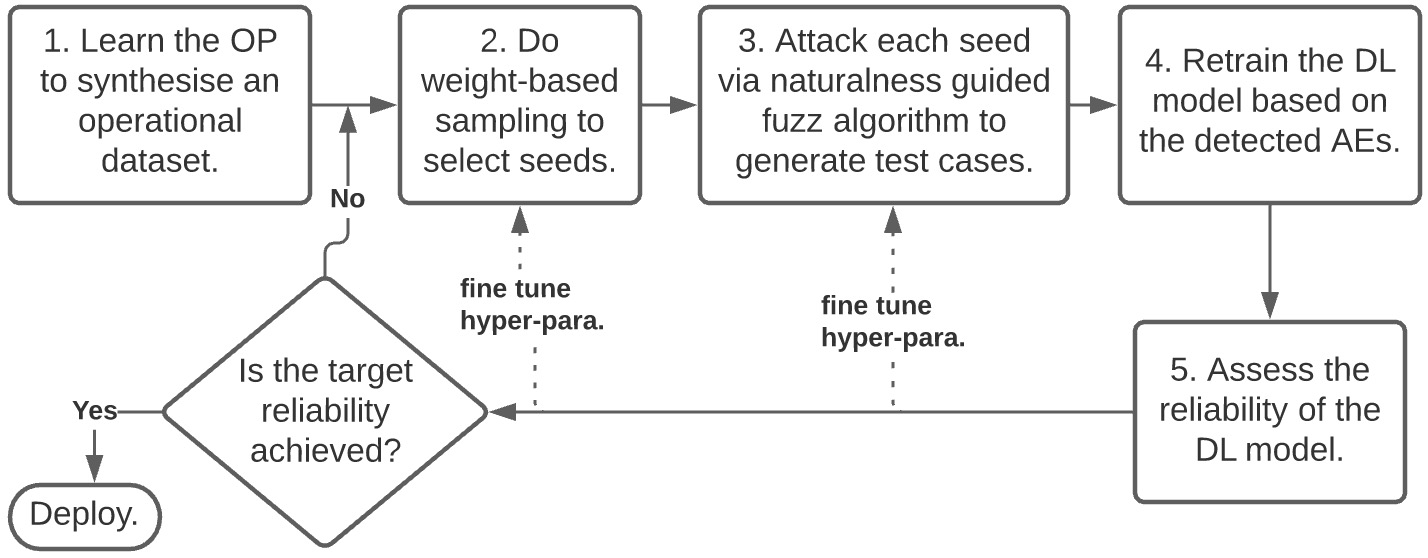}
	\caption{Workflow of the proposed solution.}
	\label{fig_framework}
\end{figure}

To achieve the aforementioned objective, we propose a five-step iterative solution, as shown in Figure~\ref{fig_framework}. Each step corresponds to a research question (RQ):
given a DL model and its specific application, we first learn the OP based on which an operational dataset is synthesised (RQ1).
Second, we design a \textit{weight-based sampling} algorithm (RQ2) to sample the ``seeds'' (of test cases)  from the operational dataset. We then implement a novel fuzzing attacking algorithm guided by naturalness (RQ3) to generate test cases around each seed. Based on the AEs detected by test cases, we retrain the DL model (RQ4). Finally, we assess the reliability of the retrained DL model (RQ5), where the result indicates how to fine tune the sampling and attacking algorithms in the next loop.
Steps 2 to 5 are repeated until the required reliability level is achieved.

Specifically, we investigate the following RQs.

(i) RQ1: How to effectively learn the OP? It is common that the future OP and the existing dataset used for training are mismatched. Training data is normally collected in a \textit{balanced} way, so that the DL model may perform well in all categories of input, especially when the OP is uncertain at the time of training. Given a DL application, in addition to conventional ways of building the OP \cite{strigini_guidelines_1997}, we envisage that techniques like high-fidelity simulation and data augmentation \cite{zhang_deeproad_2018} are essential to speed up the learning and validation of the OP.

(ii) RQ2: How to select seeds that are from high density areas on the OP and also in the ``buggy area'' of the input space? Similar questions are initially answered in \cite{guerriero_operation_2021} by using \textit{weight-based sampling} from the operational data set, where the weights are calculated based on auxiliary information that indicates which data-points are likely to cause failure. 

(iii) RQ3: Given a seed, how to generate test cases that may efficiently detect AEs considering naturalness? Although existing attacking algorithms (e.g.\ \cite{pgd_2018}) perform well in efficiently detecting AEs around seeds, constraints on naturalness (local OPs) needs to be incorporated by new fuzzing algorithms.

(iv) RQ4: How to retrain the DL model based on the detected operational AEs? 
Adversarial training provides a potential solution to the question, yet existing methods ignore the OP information.
Ideally, an enhanced adversarial training approach would consider both the OP and the detected operational AEs, while being light-weight.

(v) RQ5: How to accurately assess the delivered reliability of DL?
The reliability assessment result should not only provide a stopping rule of our testing regime, but also indicate how to fine tune the sampling and attacking algorithms in RQ2 and RQ3.
Our ongoing project \cite{zhao_safety_2020} provides a preliminary assessment model \cite{reasdl_2020}, in which OP information and robustness evidence are considered to support reliability claims.


\section{Future Work}

In our future work, we plan to develop the new DL testing tool ourlined in Figure~\ref{fig_framework} by tackling the five RQs. Although the potential methods to address each RQ are discussed, it is challenging to implement those methods and integrate them into a holistic solution.
That said, conducting comprehensive and rigorous evaluation experiments on our new approach and comparing the result to state-of-the-art is crucial, not only for the calibration of our approach but also to test (and hopefully demonstrate) its superiority, e.g. requiring significant less amount of test cases to achieve the same level of reliability.

\section*{Acknowledgment}

\includegraphics[height=8pt]{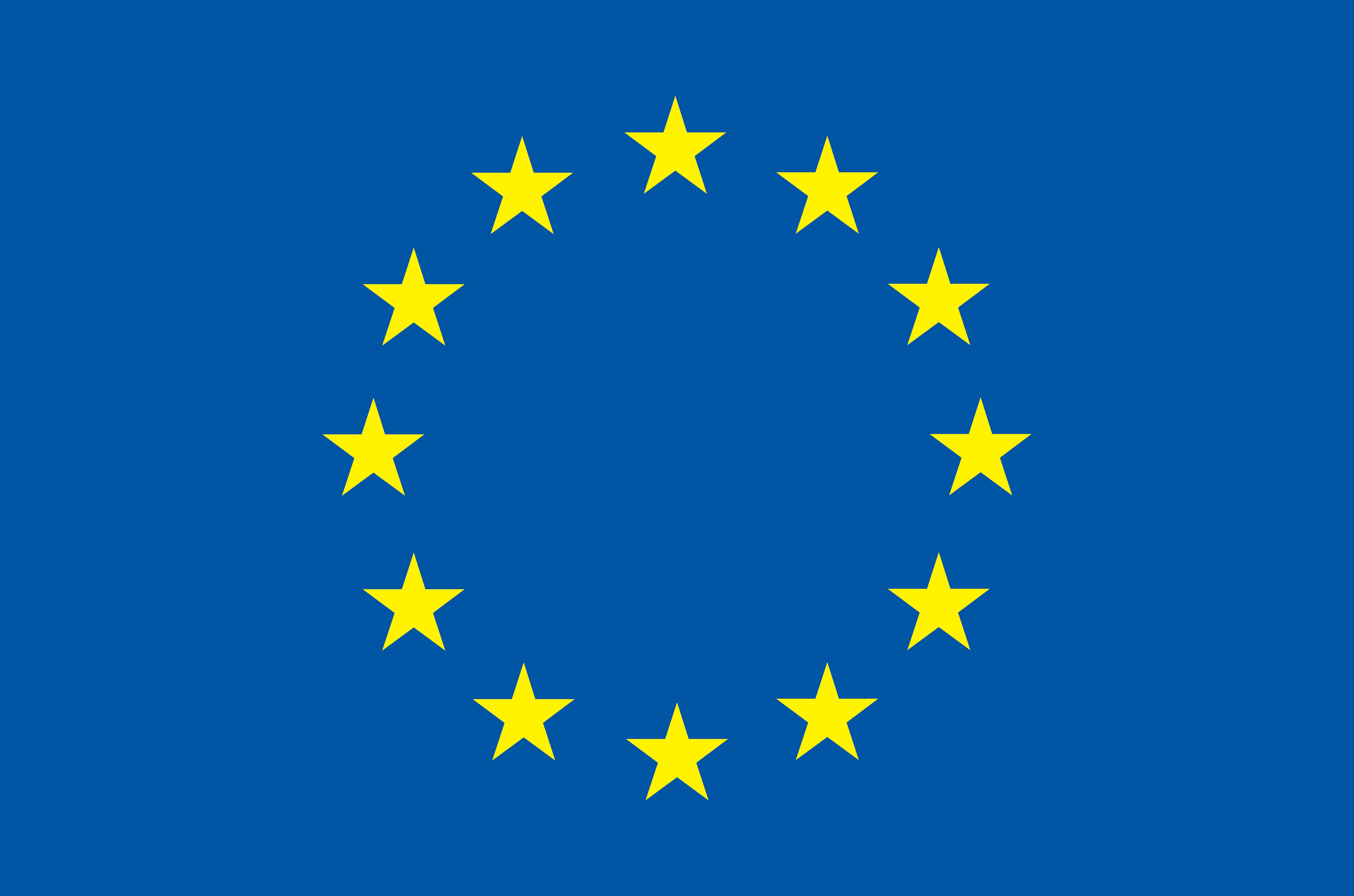} This project has received funding from the European Union’s Horizon 2020 research and innovation programme under grant agreement No 956123.
This work is partially supported by 
the UK Dstl (through the project of Safety Argument for Learning-enabled Autonomous Underwater Vehicles).
XZ’s contribution is partially supported through Fellowships at the Assuring Autonomy International Programme.
We thank Lorenzo Strigini for his insightful comments.

\bibliographystyle{IEEEtran}
\bibliography{ref}

\end{document}